# Waveguide integrated high performance magneto-optical isolators and circulators on silicon nitride platforms

**WEI YAN[1,2], YUCONG YANG[1,2], SHUYUAN LIU[1,2], YAN ZHANG[3], SHUANG XIA[1,2], TONGTONG KANG[1,2], WEIHAO YANG[1,2], JUN QIN[1,2], LONGJIANG DENG[1,2] AND LEI BI[1,2,*]**

*1National Engineering Research Center of Electromagnetic Radiation Control Materials, University of Electronic Science and Technology of China, Chengdu 610054, China*
*2State Key Laboratory of Electronic Thin-Films and Integrated Devices, University of Electronic Science and Technology of China, Chengdu, 610054, China*
*3Chongqing United Microelectronics Center, Chongqing, 410000, China*
*\*bilei@uestc.edu.cn*

**Abstract:** Optical isolators and circulators are indispensable for photonic integrated circuits (PICs). Despite of significant progress in silicon-on-insulator (SOI) platforms, integrated optical isolators and circulators have been rarely reported on silicon nitride (SiN) platforms. In this paper, we report monolithic integration of magneto-optical (MO) isolators on SiN platforms with record high performances based on standard silicon photonics foundry process and magneto-optical thin film deposition. We successfully grow high quality MO garnet thin films on SiN with large Faraday rotation up to -5900 deg/cm. We show a superior magneto-optical figure of merit (FoM) of MO/SiN waveguides compared to that of MO/SOI in an optimized device design. We demonstrate TM/TE mode broadband and narrow band optical isolators and circulators on SiN with high isolation ratio, low cross talk and low insertion loss. In particular, we observe 1 dB insertion loss and 28 dB isolation ratio in a SiN racetrack resonator-based isolator at 1570.2 nm wavelength. The low thermo-optic coefficient of SiN also ensures excellent temperature stability of the device. Our work paves the way for integration of high performance nonreciprocal photonic devices on SiN platforms.

## 1. Introduction

Optical isolators and circulators are key components in photonic integrated circuits (PICs). They prevent the reflected light from adversely affecting the lasers and the optical system [1]. Integrated optical isolators have been the missing link in semiconductor PICs. The quest for a low insertion loss, wide bandwidth, and high isolation ratio optical isolator on silicon has triggered great research interest in the past decade [2-4]. To break Lorentz reciprocity [5] in an optical isolator device, several mechanisms have been proposed based on the magneto-optical (MO) effects [6-9], the nonlinear optical effects [10, 11], and spatio-temporal modulation induced optical nonreciprocity [12-14] *etc.*. Among them, waveguide integrated MO isolators achieved significant progress recently [2, 15-17]. Integrated MO isolators and circulators have been fabricated on silicon on insulator (SOI) platforms[6, 16], showing isolation ratio up to 30 dB, insertion loss down to 5 dB for the isolators [2], and low cross talk of -30 dB at 1555 nm wavelength for the circulators [16].

However, integrated optical isolators reported so far are mostly based on SOI platforms. Devices based on silicon nitride platforms have been rarely reported. As an important material platform, SiN PICs show wide transparency window and low insertion loss, making them attractive for optical communication, data communication and optical sensing applications [18-21]. Integrated SiN photonic devices such as lasers [22], frequency comb sources [23], optical filters[24] and optical delay lines [25] have been widely reported. Integrated optical isolators and circulators on SiN are considered important for these applications[26]. The difficulty for development of such devices is due to several reasons. First, the optical nonlinearity of SiN is weaker than Si [18]. Therefore, a much higher electromagnetic field intensity is required to induce optical nonreciprocity in SiN due to optical nonlinearity. Second, compared to SOI, SiN lacks electro-optical modulation mechanisms such as plasma dispersion [27], causing difficulties for spatio-temporal modulation induced optical nonreciprocity. Third, for MO isolators based on the nonreciprocal-phase shift (NRPS) effect, the low refractive index of SiN leads to much lower NRPS compared to SOI waveguides [28], placing significant challenges for fabricating a high performance MO optical isolator device on SiN.

Despite of these challenges, progress has been made toward SiN based MO isolators. Growth of YIG/Ce: YIG thin films on SiN by pulsed laser deposition (PLD) was first reported by M. Onbasli *et al.* [29]. Faraday rotation up to -2650 $°cm^{-1}$ at around 1550 nm was demonstrated. Later, P. Pintus *et. al.* reported theoretical design of magneto-optical circulators for TM and TE modes with YIG/Ce:YIG deposited on SiN based micro-ring resonators [30]. Recently, the first experimental demonstration of SiN based TE mode optical isolators was reported by Y. Zhang *et al.* [2]. The device was fabricated by deposition of YIG/Ce:YIG thin films on the sidewall of a SiN waveguide in a micro-ring resonator. 20 dB isolation ratio and 11.5 dB insertion loss at the wavelength of 1584.9 nm was reported. The much higher insertion loss of these devices compared to their counterparts on SOI was due to the relatively low magneto-optical effects of the Ce:YIG thin film, as well as the non-ideal device geometry. Therefore, it is still uncertain whether a high-performance optical isolator can be integrated on SiN.

Here, we report high-performance integrated optical isolators and circulators on SiN PICs fabricated by standard silicon photonics foundry process and MO thin film deposition. We successfully grew $Ce_{1.4}Y_{1.6}Fe_5O_{12}$ thin films on SiN, showing Faraday rotation up to -5900 deg/cm at 1550 nm wavelength, which is about 2 times higher than $Ce_1Y_2Fe_5O_{12}$ thin films on SOI. By judiciously design the waveguide geometry, high MO waveguide figure of merit was achieved. TE/TM Mach-Zehnder Interferometer (MZI) isolators, circulators and ring resonator based isolators all show high isolation ratio, low insertion loss and low cross talk, exceeding SOI based devices. In particular, we demonstrate SiN based ring isolators with insertion loss down to 1 dB at 1570.3 nm wavelength. We believe these devices will provide pivotal functionalities of optical nonreciprocity in SiN PICs.

## 2. Device Design and Fabrication

Fig.1 (a) shows the schematics of optical isolators and circulators on SiN. The broadband and narrow band devices are based on SiN MZIs and SiN racetrack resonators, respectively [3, 31-33]. In order to reduce the device footprint, we designed the two interference arms of MZI on the same side, making the overall device shape like a paper clip. The device foot print excluding the polarization rotator is 1 mm×0.24 mm. 3 dB directional couplers are introduced at both ends of the MZI to form a four-port circulator device structure. The reciprocal phase shifter and nonreciprocal phase shifter on both arms of the MZI allow 0 (π) phase difference for the forward (backward) propagating light [34]. The nonreciprocal phase shifter is an MO waveguide consisting of Ce:YIG/YIG thin films deposited on top of SiN channel waveguides, providing NRPS for TM polarized light [16]. Compared to our previous design of Ce:YIG/YIG thin films on the sidewalls of SiN waveguides [2], current devices showed higher MO/SiN waveguide figure of merit, as will be discussed later. For TE mode isolators, TE-TM polarization rotators (PRs) are inserted at both ends of the TM isolators [16, 31, 35]. The polarization rotator was based on the mode evolution mechanism to rotate the fundamental TE mode into the fundamental TM mode [36, 37]. For the narrow-band optical isolators based on racetrack resonators, the MO waveguide is arranged outside the coupling region between the ring waveguide and the bus waveguide. Optical isolation is achieved by the resonance frequency splitting due to NRPS in the resonator cavity [2, 3].

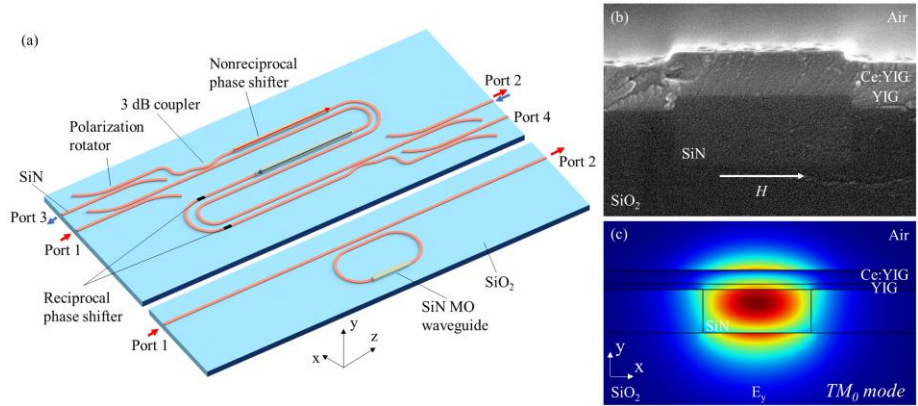

Fig. 1. (a) Schematics of integrated magneto-optical isolators on SiN. (b) SEM image of the cross section of a fabricated MO/SiN waveguide. (c) Simulated $E_y$ field distribution of the fundamental TM mode of the MO/SiN waveguide.

Fig. 1 (b) shows the cross-sectional scanning electron microscope (SEM) image of a fabricated MO/SiN waveguide. The device consists of a planar Ce:YIG (150 nm)/YIG (50 nm) thin film stack deposited on SiN, forming a strip-loaded waveguide structure. This structure prevented deposition of the MO thin films on the sidewalls of SiN, which was considered detrimental to the waveguide FoM for TM mode nonreciprocal phase shifters [4]. Details of the fabrication process can be found in Methods. An in-plane external magnetic field (1000 Gs) was applied perpendicular to the propagation direction of light to saturate the magnetization of the MO thin films. Fig. 1 (c) shows the simulated $E_y$ field distribution of the fundamental TM mode. From the figure we can see that the intensity of the modal profile is obviously discontinuous at the interface of different materials along the y direction due to the boundary conditions of electromagnetic fields [38].

Next, the isolator device structure is designed aiming to achieve the best MO/SiN

waveguide FoM. Different from SOI with a standard film thickness of 220 nm, the thickness of SiN waveguides can be optimized, together with the MO film thicknesses and MO effects. The influence of the material's MO effect and waveguide geometry on the figure of merit of the MO/SiN waveguides was firstly studied. The MO waveguide FoM ($Q_{MOWG}$) is defined as the ratio between its NRPS and propagation loss of the fundamental TM mode:

$$Q_{MOWG} = \frac{\Delta\beta}{\alpha_{MOWG}} \tag{1}$$

$$\Delta\beta = \frac{2\beta}{\omega\varepsilon_0 N} \iint \frac{\gamma}{n_0^4} H_x \partial_y H_x \, dxdy \tag{2}$$

$$\alpha_{MOWG} = \Gamma_{YIG}\alpha_{YIG} + \Gamma_{SiN}\alpha_{SiN} \tag{3}$$

Where $\Delta\beta$ and $\beta$ are the NRPS and propagation constant of the fundamental TM mode, respectively [39]. $\omega$ is the angular frequency, $\gamma$ is the off diagonal component of the permittivity tensor of the MO material [39], $\varepsilon_0$ is the vacuum dielectric constant, $N$ is the power flux along z direction, $n_0$ is the refractive index of the magneto-optical material, $H_x$ is the magnetic field along x direction, $\Gamma_{YIG}$ and $\Gamma_{SiN}$ are the confinement factors in Ce:YIG/YIG and SiN waveguide cores respectively [40]. From equation (1) to (3), the key to achieve a high MO waveguide FoM is to obtain a high NRPS $\Delta\beta$ while keeping the propagation loss $\alpha_{MOWG}$ low. $\Delta\beta$ is directly proportional to $\gamma$, which increases with Ce ion concentration in $Ce_xY_{3-x}Fe_5O_{12}$ [41]. Fig. 2 (a) shows the room temperature Faraday rotation of $Ce_xY_{3-x}Fe_5O_{12}$ thin films at 1550 nm wavelength with different Ce concentrations deposited on Si substrates. The Faraday rotation increased with Ce concentrations. A high Faraday rotation up to -5900 °cm$^{-1}$ was observed in $Ce_{1.4}Y_{1.6}Fe_5O_{12}$ thin films, which was about 100% higher than $Ce_1Y_2Fe_5O_{12}$ in previous reports[41]. On the other hand, the waveguide propagation loss $\alpha_{MOWG}$ can be calculated based on the propagation loss of Ce:YIG/YIG, SiN materials and the modal profile, assuming 0 loss from the $SiO_2$ claddings. In the simulations, we set $\alpha_{YIG}$ =110 dB/cm [42] and $\alpha_{SiN}$ =0.5 dB/cm [43]. The thicknesses of Ce: YIG and YIG were set as 100 nm and 50 nm respectively. The YIG layer was set with a Faraday rotation of 500 °cm$^{-1}$ [44]. Based on these material parameters, we simulated the relationship between the FoM of the

MO/SiN waveguide and the width/thickness of the SiN waveguide core, as shown in Fig. 2 (b). The results show that when the width of the SiN waveguide varies from 800 nm to 1400 nm, the corresponding optimal SiN waveguide thicknesses are between 360 nm and 440 nm. The curves show a rather flat top, indicating good fabrication tolerance. Thinner SiN waveguide thickness is beneficial for a larger FoM. In order to avoid the influence of waveguide thickness deviation caused by the fabrication process, we choose the center area within the flat top. For SiN waveguide thickness of 400 nm and width of 1000 nm, a FoM of 1.043 rad/dB can be achieved.

Next, as shown Fig. 2 (c), we fixed the SiN waveguide core dimension as 400 nm × 1000 nm, the YIG layer thickness as 50 nm, and calculated the MO/SiN waveguide loss as a function of the Ce:YIG layer thickness and Faraday rotation for achieving NRPS=$\pi/2$, using the measured Faraday rotation values in Fig. 2 (a). The propagation loss of MO/SiN and MO/SOI waveguides is also compared using the same Ce:YIG/YIG materials and waveguide cross-section structures, as shown in Fig. 1 (c). In these simulations, the SOI waveguide core dimension was set as 220 nm × 500 nm [8]. The propagation loss of SOI waveguides was set as 3 dB/cm [45]. Two differences can be immediately observed between SiN and SOI based devices. First, the propagation loss of the MO/SOI waveguide monotonically decreases with increasing Ce:YIG film thickness, whereas the MO/SiN waveguide shows the lowest loss for Ce:YIG film thickness around 150 nm. This is because of the high refractive index of Si (n=3.48) compared to Ce: YIG (n=2.3), which requires thicker Ce:YIG films for a larger NRPS. Whereas for relatively low index SiN (n=2.0), a thinner Ce:YIG is required. This difference makes SiN advantageous compared to SOI for deposited garnet films. Because the thickness of deposited Ce:YIG/YIG has to be kept thin to prevent cracks due to the thermal mismatch between the garnet materials and the substrate [41]. Second, lower loss is always observed in MO/SiN waveguides for the simulated parameter range. This is essentially due to a smaller confinement factor in the MO layers of the MO/SiN waveguide, as well as much lower loss of the SiN waveguide itself compared to SOI. In terms of device footprint, Fig. 2 (d) shows the length of each MO waveguide in the cases corresponding to Fig. 2 (c). The MO waveguide length monotonically decreases when increasing the Ce:YIG film thickness, leading to larger NRPS. The MO/SOI waveguide is always about 2 times shorter compared to the MO/SiN waveguide due to the higher NRPS. Therefore, a high Faraday rotation in $Ce_{1.4}Y_{1.6}Fe_5O_{12}$ up to -5900° is crucial to bring the MO/SiN waveguide length down to around 600 μm, which is comparable to the MO/SOI waveguide length using $Ce_1Y_2Fe_5O_{12}$ with Faraday rotation of -3800°.

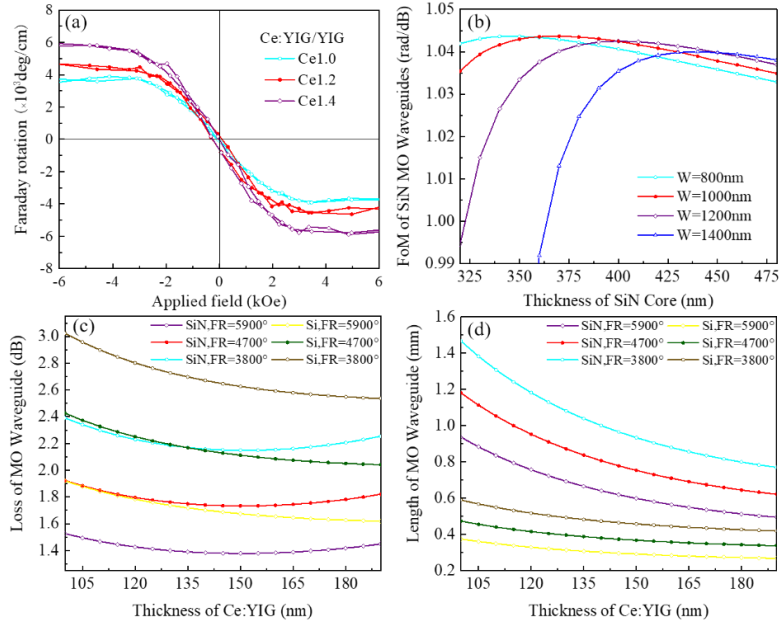

Fig. 2. (a) Faraday rotation of Ce:YIG thin films with different cerium concentrations deposited on silicon substrates. (b) The relationship between the FoM of the MO/SiN waveguide and the SiN core thickness under different SiN core widths. (c) The MO waveguide loss as a function of Ce:YIG layer thickness, for different Faraday rotation values of Ce:YIG. (d) The corresponding lengths of the MO/SiN waveguide and the MO/SOI waveguide for different Faraday rotation values of Ce:YIG.

## 3. Device Characterization and Discussion

The fabricated devices were characterized on a polarization maintaining, fiber-butt-coupled system as detailed in Methods. Fig. 3 (a) shows the transmission spectra of the TM mode SiN optical isolator, together with a reference silicon nitride waveguide on the same chip. At 1555 nm wavelength, the maximum isolation ratio reached 32 dB, with an insertion loss of 2.3 dB. The 20 dB and 10 dB isolation bandwidth of this device are 4 nm and 13 nm respectively. The device bandwidth can be further increased by reducing the RPS waveguide length [46]. Across the entire 10 dB isolation bandwidth, the device shows insertion loss of 2.3-3 dB. To the best of our knowledge, this represents the lowest insertion loss measured in a broadband on-chip optical isolator reported so far. The device insertion loss is partly due to MO/SiN waveguide propagation loss (2 dB) and the MO/SiN and SiN waveguide junction loss (0.3 dB), which can be further reduced with material improvements. Fig. 3 (b) shows the transmission spectra of the TE mode optical isolator. A maximum isolation ratio of 30 dB, insertion loss of 3 dB and 10 dB isolation bandwidth of 16 nm are observed at 1558 nm wavelength. The extra loss (0.7 dB) compared to the TM isolator is mainly due to the two polarization rotators, which can also be improved in future designs. The result indicates that Faraday rotation of $Ce_{1.4}Y_{1.6}Fe_5O_{12}$ deposited on SiN waveguides also reaches up to -5900 deg/cm, consistent with thin film characterization results in Fig. 2 (a). The transmission spectra of the TM optical circulator are characterized in Fig. 3 (c). Based on the circulator direction: Port 1 → Port 2 → Port 3 → Port 4, the corresponding spectra are shown with two opposite directions of Port 2 → Port 1 and Port 3 → Port 2. The crosstalk between two adjacent output ports ranges from -15.7 dB to -32 dB at 1555 nm wavelength. The insertion loss of the device is between 2.3 dB and 3.8 dB for different output ports, when compared to a SiN reference TM waveguide at the same wavelength. Fig. 3 (d) shows the transmission spectra of a TE optical circulator, crosstalk from -25.3 dB to -30 dB and insertion loss between 3 dB and 4.4 dB are achieved at 1558 nm wavelength. Complete performance report of both TM and TE SiN optical circulators normalized to SiN reference waveguides is shown in Table 1.

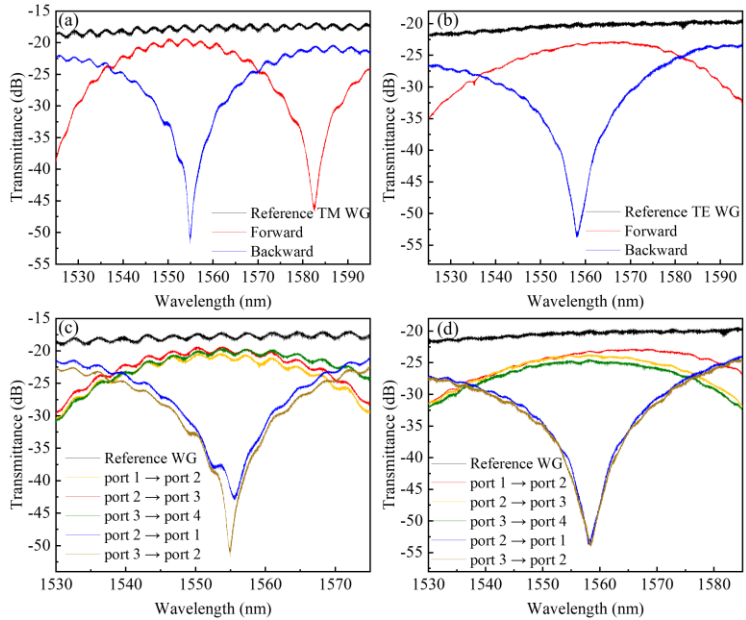

Fig. 3. (a)(b) The transmission spectra of the broadband TM and TE mode SiN optical isolators.
(c)(d) The transmission spectra of the broadband TM and TE mode SiN optical circulators.

**Table 1. Optical transmittance for TM (TE) SiN optical circulators between different ports (normalized to a TM (TE) SiN reference waveguide)**

| Input \ Output | TM Polarization (dB) @ 1555 nm | | | | TE Polarization (dB) @ 1558 nm | | | |
|---|---|---|---|---|---|---|---|---|
| | Port 1 | Port 2 | Port 3 | Port 4 | Port 1 | Port 2 | Port 3 | Port 4 |
| Port 1 | - | -3.7 | - | -19.4 | - | -3 | - | -28.4 |
| Port 2 | -25 | - | -2.3 | - | -33.5 | - | -3.5 | - |
| Port 3 | - | -34.4 | - | -2.4 | - | -33.4 | - | -4.4 |
| Port 4 | -3.8 | - | -23.7 | - | -3.1 | - | -28.4 | - |

The temperature stability of the isolator devices is also characterized. The device temperature was controlled at 20 °C, 45 °C and 70 °C by a heater stage. Fig. 4 shows the transmission spectra of both TM and TE broadband SiN optical isolators under the above-mentioned temperatures. The spectra shifted with increasing temperature because the refractive index of SiN and Ce:YIG, as well as the Faraday rotation coefficient of Ce:YIG are temperature dependent [46]. To analyze this, the thermo-optic coefficient of SiN ($dn/dT = 4.7 \times 10^{-5}$/K [47]), the thermo-optic coefficient of Ce:YIG ($dn/dT = 9.1 \times 10^{-5}$/K) and the Faraday rotation temperature coefficient of Ce:YIG ($d\theta_F/dT = 44°$/cm/K [46]) are considered. For the backward direction, the temperature dependence of phase differences $\theta_{NRPS}(T)$ and $\theta_{RPS}(T)$ partially cancelled with each other, because the refractive index increases and the MO effect becomes weaker with increasing temperature. Whereas for the forward direction, the temperature influence of phase differences $\theta_{NRPS}(T)$ and $\theta_{RPS}(T)$ added up to each other, and the corresponding curve changed more obviously [46]. In general, SiN isolators maintained a relatively stable operation wavelength within the temperature range from 20 °C to 70 °C, showing a small wavelength drift within 4 nm. This value is smaller compared to a drift over 6 nm for SOI devices from 20 °C to 60 °C [46], due to the larger thermo-optic coefficient of Si ($1.8 \times 10^{-4}$/K) [48].

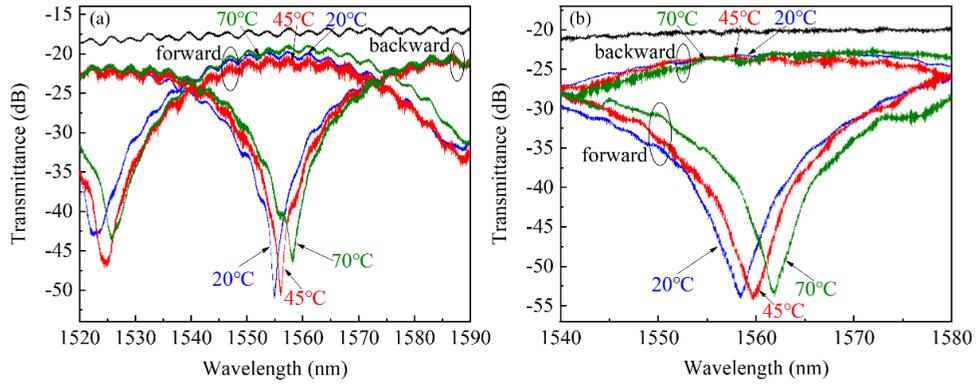

Fig. 4. Transmission spectra of broadband SiN optical isolators under different device temperatures of 20 °C, 45 °C and 70 °C respectively for (a) TM and (b) TE polarizations.

Fig. 5 (a) shows the optical microscope image of a TM mode optical isolator based on a SiN racetrack resonator. A record low insertion loss of 1 dB is observed at 1570.28 nm wavelength, as shown in Fig. 5 (b). The isolation ratio reaches 28 dB at resonance. The high performance is essentially due to the strong magneto-optical effect of $Ce_{1.4}Y_{1.6}Fe_5O_{12}$ and the optimized MO/SiN waveguide design, leading to 120 pm resonance peak shift, which is 6 times higher than our previous report [2]. The quality factor (loaded Q) of the device at around critical coupling is calculated to be 13000. The free spectral range (FSR) is 1.5 nm. This quality factor is also 2 times higher than that of SOI based devices [3].

To benchmark the performance of our devices, Table 2 compares several recently reported MO isolators on SOI and SiN platforms. We observe higher performance in SiN based MO isolators reported in this work. In the future, the device performance can be further improved. Currently, 85% of the insertion loss is due to the propagation loss of the MO material, by further improving the MO material figure of merit, the insertion loss can be improved. Reducing the YIG seed layer thickness or using a top seed layer can also lead to a much higher waveguide FoM by increasing coupling of the waveguide mode from the waveguide core into the Ce: YIG layer, which has already been reported in several material systems recently [42, 49, 50].

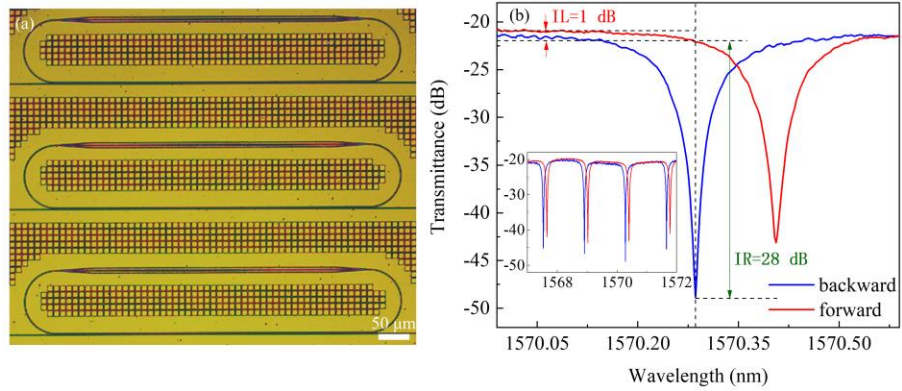

Fig. 5 (a) Optical microscope image of the TM mode optical isolator based on SiN racetrack resonators. (b) The transmission spectra of the isolators.

Table 2. Comparison of MO isolator device performances near the C-band

| Device type | Isolation Ratio (dB) | Insertion loss (dB) | Size (mm$^2$) | Monolithic / Bonding | Polarization | 10 dB Bandwidth (nm) | Ref. |
|---|---|---|---|---|---|---|---|
| **SiN MZI** | **32** | **2.3** | **1.0× 0.24** | **Monolithic** | **TM** | **13** | **This work** |
| **SiN MZI** | **30** | **3** | **1.95× 0.24** | **Monolithic** | **TE** | **16** | **This work** |
| Si MZI | 30 | 14 | N.A. | Bonding | TE | 14 | [16] |
| Si MZI | 25 | 8 | 4×4 | Bonding | TM | ~1 | [51] |
| Si MZI | 30 | 5 | 0.94× 0.33 | Monolithic | TM | 9 | [2] |
| SiN MZI | 18 | 10 | 3.2× 1.0 | Monolithic | TE | 3 | [2] |
| **SiN Ring** | **28** | **1** | **0.6× 0.1** | **Monolithic** | **TM** | **N.A.** | **This work** |
| SiN Ring | 20 | 11.5 | 0.3× 0.45 | Monolithic | TE | N.A. | [2] |
| Si Ring | 32 | 2.3 | 0.07× 0.07 | Bonding | TM | N.A. | [32] |

## 4. Conclusion

In summary, high performance, monolithically integrated optical isolators and circulators for both TE and TM polarizations on SiN have been experimentally demonstrated in this report. The devices are fabricated using standard silicon photonics foundry processes for SiN PICs and PLD for the MO thin films. $Ce_{1.4}Y_{1.6}Fe_5O_{12}$ thin films are deposited on SiN with high Faraday rotation up to -5900°/cm at 1550 nm. For thin garnet films below 200 nm, the MO/SiN waveguides show higher waveguide FoM compared to MO/SOI in an optimized design. High performance broadband TM(TE) isolators and circulators with 32 dB(30 dB) isolation ratio, 2.3 dB(3 dB) insertion loss and -32 dB(-30 dB) minimum crosstalk are demonstrated, surpassing 220 nm SOI based devices. Narrow band optical isolators based on SiN racetrack resonators show record low insertion loss down to 1 dB and isolation ratio up to 28 dB. Thanks to the low thermo-optic coefficient of SiN, the devices also show small wavelength drift within 4 nm in the tested temperature range of 20 °C to 70 °C. Our work represents an important step towards integration of nonreciprocal photonic devices on SiN PICs, paving the way for a variety of applications in telecommunication [23], data communication [52], frequency comb generation and on-chip LIDAR systems [53].

**Methods**

*Device fabrication*

The proposed devices were first manufactured in a silicon photonics foundry. The SiN waveguide was prepared on a $SiO_2$ bottom cladding by low pressure chemical vapor deposition (LPCVD) and reactive ion etching (RIE). A chemical mechanical polishing (CMP) step was utilized to planarize the top $SiO_2$ cladding, followed by RIE to expose the SiN waveguide core at the MO waveguide section. Then, 50 nm YIG and 150 nm Ce:YIG thin films were deposited by PLD [2, 3, 41].

*Device characterization*

The devices were characterized on a polarization maintaining, fiber-butt-coupled system. During the test, an external in-plane magnetic field of 1000 Oe was applied in the direction perpendicular to the propagation light in MO/SiN waveguides to saturate the garnet films. A tunable laser (Keysight 81960A) provided a continuous wave spectrum with a wavelength from 1520 nm to 1610 nm, and the input optical power was 8 dBm. A free-space polarization control bench was used to obtain TE or TM polarized light. The light was coupled into the device under test by means of butt coupling through a lens tipped PM fiber. The output signal was then coupled into another fiber connected to an optical power sensor (Agilent 81636B). The wavelength detection interval was set as 2 pm. During the backward transmission test, the direction of the applied magnetic field remained unchanged, and the connection status of the input and output fiber with the laser and the sensor respectively were switched through a 2 by 2 fiber optical switch. For the temperature dependence test, we heated the substrate holder and adjusted the heating device to obtain a desired temperature via a temperature sensor.

## Disclosures

The authors declare no conflicts of interest.

## References

1. P. K, "External optical feedback phenomena in semiconductor lasers," IEEE Journal of Selected Topics in Quantum Electronics **1**, 480-489 (1995).
2. D. Q. Zhang Y, Wang C, et al., "Monolithic integration of broadband optical isolators for polarization-diverse silicon photonics," Optica **6**, 473-478 (2019).
3. H. J. Bi L, Jiang P, et al., "On-chip optical isolation in monolithically integrated non-reciprocal optical resonators," Nature Photonics **5**, 758 (2011).
4. W. C. Du Q, Zhang Y, et al., "Monolithic on-chip magneto-optical isolator with 3 dB insertion loss and 40 dB isolation ratio," ACS Photonics **12**, 5010-5016 (2018).
5. P. A. Jalas D, Eich M, et al., "What is—and what is not—an optical isolator," Nature Photonics **7**, 579-582 (2013).
6. D. P. Zhang C, Stadler B J H, et al., "Monolithically-integrated TE-mode 1D silicon-on-insulator isolators using seedlayer-free garnet," Scientific reports **7**, 1-8 (2017).
7. I. T. Espinola R L, Tsai M C, et al., "Magneto-optical nonreciprocal phase shift in garnet/silicon-on-insulator waveguides," Optics letters **29**, 941-943 (2004).
8. M. K. Shoji Y, Mizumoto T, "Optical nonreciprocal devices based on magneto-optical phase shift in silicon photonics," Journal of Optics **18**, 013001 (2015).
9. S. V. Karki D, Pollick A, et al., "Broadband Bias-Magnet-Free On-Chip Optical Isolators With Integrated Thin Film Polarizers," Journal of Lightwave Technology **38**, 827-833 (2019).
10. W. J. Fan L, Varghese L T, et al., "An all-silicon passive optical diode," Science **335**, 447-450 (2012).
11. S. Y. Wang J, Fan S, "Non-reciprocal polarization rotation using dynamic refractive index modulation," Optics Express **28**, 11974-11982 (2020).
12. O. N. T. Kittlaus E A, Kharel P, et al., "Non-reciprocal interband Brillouin modulation," Nature Photonics **12**, 613-619 (2018).
13. K. S. Sohn D B, Bahl G, "Time-reversal symmetry breaking with acoustic pumping of nanophotonic circuits," Nature Photonics **12**, 91-97 (2018).
14. N. Dostart, Yossef Ehrlichman, Cale M. Gentry and Miloš Popović, "Energy-efficient active integrated photonic isolators using electrically driven acoustic waves," arXiv, 1811.01052 (2018).
15. P. P. Huang D, Peters J, et al, "Widely Tunable Ce: YIG on Si Microring Isolators for TE Mode Operation," in *2018 IEEE 15th International Conference on Group IV Photonics (GFP)*, (IEEE, 2018), 1-2.
16. H. D. Pintus P, Morton P A, et al., "Broadband TE optical isolators and circulators in silicon photonics through Ce: YIG bonding," Journal of Lightwave Technology **37**, 1463-1473 (2019).
17. S. Y. Mizumoto T, "On-chip Optical Isolators," in *Optical Fiber Communication Conference*, (Optical Society of America, 2020), T3B. 1.
18. S. A. Z. Baets R, Clemmen S, et al., "Silicon Photonics: silicon nitride versus silicon-on-insulator," in *Optical Fiber Communication Conference*, (Optical Society of America, 2016), Th3J. 1.
19. D. N. Melchiorri M, Sbrana F, et al., "Propagation losses of silicon nitride waveguides in the near-infrared range," Applied Physics Letters **86**, 121111 (2005).
20. M. G. Muñoz P, Bru L A, et al., "Silicon nitride photonic integration platforms for visible, near-infrared and

mid-infrared applications," Sensors **17**, 2088 (2017).
21. H. R. Blumenthal D J, Geuzebroek D, et al., "Silicon nitride in silicon photonics," Proceedings of the IEEE **106**, 2209-2231 (2018).
22. O. R. M. Fan Y, Roeloffzen C G, et al., "290 Hz intrinsic linewidth from an integrated optical chip-based widely tunable InP-Si3N4 hybrid laser," in *CLEO: QELS_Fundamental Science*, (Optical Society of America, 2017), JTh5C. 9.
23. L. M. Gaeta A L, Kippenberg T J., "Photonic-chip-based frequency combs," Nature Photonics **3**, 158-169 (2019).
24. R. C. G. H. Zhuang L, Hoekman M, et al., "Programmable photonic signal processor chip for radiofrequency applications," Optica **2**, 854-859 (2015).
25. Z. L. Roeloffzen C G H, Heideman R G, et al., "Ring resonator-based tunable optical delay line in LPCVD waveguide technology," in *Proc. 9th IEEE/LEOS Symp*, (Benelux, 2005), 79-82.
26. P. P. Bowers J E, Heck M J R, et al., "Integrated optical circulators and isolators on a ultra-low-loss silicon nitride platform," in *2013 IEEE Photonics Society Summer Topical Meeting Series*, (IEEE, 2013), 201-202.
27. R. G. T. Gardes F Y, Knights A P, et al., "Evolution of optical modulation using majority carrier plasma dispersion effect in SOI," in *Silicon Photonics III*, (International Society for Optics and Photonics, 2008), 6898: 68980C.
28. C. J. Zhou H, Song J, et al. , "Analytical calculation of nonreciprocal phase shifts and comparison analysis of enhanced magneto-optical waveguides on SOI platform," Optics express **8**, 8256-8269 (2012).
29. G. T. Onbasli M C, Sun X, et al., "Integration of bulk-quality thin film magneto-optical cerium-doped yttrium iron garnet on silicon nitride photonic substrates," Optics express **22**, 25183-15192 (2014).
30. P. Pintus, Fabrizio Di Pasquale, and John E. Bowers, "Integrated TE and TM optical circulators on ultra-low-loss silicon nitride platform," Optics express **21**, 5041-5052 (2013).
31. F. A. Shoji Y, Mizumoto T, "Silicon waveguide optical isolator operating for TE mode input light," IEEE Journal of Selected Topics in Quantum Electronics **22**, 264-270 (2016).
32. P. P. D. Huang, C. Zhang, Y. Shoji, T. Mizumoto, and J. E. Bowers, "Silicon microring isolator with large optical isolation and low loss," in *Optical Fiber Communication Conference*, (Optical Society of America, 2016), Th1K. 2.
33. S. Y. Shoji Y, Mizumoto T, "Silicon Mach–Zehnder interferometer optical isolator having 8 nm bandwidth for over 20 dB isolation," Japanese Journal of Applied Physics **53**, 022202 (2014).
34. K. S. Ghosh S, Shoji Y, et al, "Compact Mach–Zehnder interferometer Ce: YIG/SOI optical isolators," IEEE Photonics Technology Letters **24**, 1653-1656 (2012).
35. P. P. Huang D, Bowers J E., "Towards heterogeneous integration of optical isolators and circulators with lasers on silicon," Optical Materials Express **9**, 2471-2483 (2018).
36. L. Z. Yin Y, Dai D, "Ultra-broadband polarization splitter-rotator based on the mode evolution in a dual-core adiabatic taper," Journal of Lightwave Technology **35**, 2227-2233 (2017).
37. A. M. Z. Sun X, Aitchison J S, et al., "Polarization rotator based on augmented low-index-guiding waveguide on silicon nitride/silicon-on-insulator platform," Optics Letters **14**, 3229-3232 (2016).
38. M. H, *Boundary Conditions for Electromagnetic Fields* (Microwave Techniques, 1971).
39. Y. H, "Calculation of nonreciprocal phase shift in magneto-optic waveguides with Ce: YIG layer," Optical Materials **31**, 189-192 (2008).
40. D. P. Sun R, Feng N, et al, "Horizontal single and multiple slot waveguides: optical transmission at λ= 1550 nm," Optics express **15**, 17967-17972 (2007).
41. W. C. T. Zhang Y, Liang X, et al., "Enhanced magneto-optical effect in Y1.5Ce1.5Fe5O12 thin films deposited on silicon by pulsed laser deposition," Journal of Alloys and Compounds **703**, 591-599 (2017).
42. D. Q. Zhang Y, Wang C, et al., "Dysprosium substituted Ce: YIG thin films with perpendicular magnetic anisotropy for silicon integrated optical isolator applications," APL Materials **8**, 081119 (2019).
43. H. M. J. R. Bauters J F, John D D, et al., "Ultra-low-loss single-mode Si3N4 waveguides with 0.7 dB/m propagation loss," in *European Conference and Exposition on Optical Communications*, (Optical Society of America, 2011), Th. 12.
44. O. M. C. Goto T, Ross C A, "Magneto-optical properties of cerium substituted yttrium iron garnet films with reduced thermal budget for monolithic photonic integrated circuits," Optics express **20**, 28507-28517 (2012).
45. D. P. Teng J, Bogaerts W, et al, "Athermal Silicon-on-insulator ring resonators by overlaying a polymer cladding on narrowed waveguides," Optics express **17**, 14627-14633 (2009).
46. M. T. Shoji Y, "Silicon waveguide optical isolator with directly bonded magneto-optical garnet," Applied Sciences **3**, 609 (2019).
47. G. I. B. Zanatta A R, "The thermo optic coefficient of amorphous SiN films in the near-infrared and visible regions and its experimental determination," Applied Physics Express **4**, 042402 (2013).
48. S. C. Komma J, Hofmann G, et al., "Thermo-optic coefficient of silicon at 1550 nm and cryogenic temperatures," Applied Physics Letters **101**, 041905 (2012).
49. G. T. E. Srinivasan K, Stadler B J H., "Seed-layer free cerium-doped terbium iron garnet on non-garnet substrates for photonic isolators," in *CLEO: Science and Innovations*, (Optical Society of America, 2018), SW4I. 5.
50. D. Q. Sun X Y, Goto T, et al., "Single-step deposition of cerium-substituted yttrium iron garnet for monolithic on-chip optical isolation," Acs Photonics **7**, 856-863 (2015).
51. K. S. Ghosh S, Van Roy W, et al, "Ce: YIG/Silicon-on-Insulator waveguide optical isolator realized by

adhesive bonding," Optics express **20**, 1839-1848 (2012).
52.     T. M. R. T. Cheung S S, "Silicon Nitride (Si 3 N 4)(De-) Multiplexers for 1-μm CWDM Optical Interconnects," Journal of Lightwave Technology **13**, 3404-3413 (2020).
53.     S. J. Yang K Y, Cotrufo M, et al., "Inverse-designed non-reciprocal pulse router for chip-based LiDAR," Nature Photonics **6**, 369-374 (2020).